\documentclass[british,english]{elsarticle}
\usepackage[T1]{fontenc}
\usepackage[latin9]{inputenc}
\usepackage{babel}
\usepackage{amsmath}
\usepackage{amssymb}
\usepackage{graphicx}
\usepackage[unicode=true,
 bookmarks=false,
 breaklinks=false,pdfborder={0 0 1},backref=section,colorlinks=false]
 {hyperref}
\usepackage{breakurl}

\makeatletter

\providecommand{\tabularnewline}{\\}

\usepackage{babel}
\usepackage{babel}
\journal{Physica A}

\makeatother

\begin{document}

\title{Fluctuation analysis of high frequency electric power load in the
Czech Republic\tnoteref{thanks}}

\tnotetext[thanks]{We would like to thank to I. Jex and J. Tolar for support of this
analysis.}

\author{Ji\v{r}í Kracík\corref{coauth}}

\address{Charles University in Prague, Faculty of Social Sciences, Institute
of Economic Studies, Opletalova 26. CZ-11000 Prague 1, Czech Republic}

\ead{nyrlem.astro@seznam.cz}

\author{Hynek Lavi\v{c}ka\fnref{auth}}

\cortext[auth]{ Corresponding author}

\address{Czech Technical University in Prague, Faculty of Nuclear Sciences
and Physical Engineering, Department of Physics, B\v{r}ehová 7, CZ-11519
Prague 1, Czech Republic}

\address{Bogolyubov Laboratory of Theoretical Physics, Joint Institute of
Nuclear Research, RU-141980 Dubna, Russia}

\ead{Tel.: +420 2 2435 8352 Fax.: +420 2 2232 0861 hynek.lavicka@fjfi.cvut.cz}
\begin{abstract}
We analyze the electric power load in the Czech Republic (CR) which
exhibits a seasonality as well as other oscillations typical for European
countries. Moreover, we detect 1/f noise property of electrical power
load with extra additional peaks that allows to separate it into a
deterministic and stochastic part. We then focus on the analysis of
the stochastic part using improved Multi-fractal Detrended Fluctuation
Analysis method (MFDFA) to investigate power load datasets with a
minute resolution. Extracting the noise part of the signal by using
Fourier transform allows us to apply this method to obtain the fluctuation
function and to estimate the generalized Hurst exponent together with
the correlated Hurst exponent, its improvement for the non-Gaussian
datasets. The results exhibit a strong presence of persistent \foreignlanguage{british}{behaviour}
and the dataset is characterized by a non-Gaussian skewed distribution.
There are also indications for the presence of the probability distribution
that has heavier tail than the Gaussian distribution.\end{abstract}
\begin{keyword}
MFDFA, electric power load, Hurst exponent, persistent process, 1/f
noise, non-Gaussian distribution
\end{keyword}
\maketitle

\section{Introduction}

The responsibility for the safe and reliable operation is one of the
basic duties of the national Transmission System Operator (TSO). The
gradual liberalization of the European electricity market led to a
necessity of the integration of mutually uncoordinated transmission
systems. The enhancements of these transmission systems are very intensive
in terms of both the time as well as capital investments and due to
this the current energy networks are reaching their technical limits.
That is mostly obvious in case of a massive increase of the offshore
wind power plant installations located in the distant parts, hundreds
of kilometers far from the end consumer. The electricity, which cannot
pass through the under-dimensioned transmission lines or so called
congestions, flows through the surrounding system which must accommodate
these unscheduled flows. Unfortunately, the market with electricity
and its mechanisms do not reflect this fact. In our work, we analyze
high frequency data of electricity consumption in the Czech Republic
and we also determine the degree of the uncertainty of the behavior
of the consumers.

The analysis of the electricity prices and loads has been discussed
by R. Weron \citep{key02}. He stated that the electricity loads,
which are non-stationary time series, are combinations of both the
trends and the periodic cycles with a random component. It is known
from literature that electricity loads are correlated with the weather
(e.g., the temperature, see \citep{key02,Peirson1994235,Lee2011896})
as well as with socio-economical changes and processes.

The first method (R/S method) for a non-stationary time series analysis
was invented by H.E. Hurst \citep{key29}. Since its introduction
the method has been tested on various datasets and also implemented
very effectively on computer \citep{key08,key16,key17,key20}. The
method estimates the Hurst exponent of dataset that is related to
the exponent of the autocorrelation function from the theory of fractional
Brownian motion \citep{key32,key24}. A modern alternative of the
Hurst exponent estimation for series with local trends is the Detrended
Fluctuation Analysis (DFA) which was introduced in \citep{key14,key15}
and used for economy datasets \citep{key03}, heart rate dynamics
\citep{key23,key25}, DNA sequences \citep{key26,key15,key14}, long-time
weather records \citep{key13}, electricity prices time series \citep{key16,key17}
and wind speed records \citep{key37}. Recently, the DFA was improved
to quantify the fluctuation function of datasets using different metrics
\citep{key06,key07}. The MFDFA is able to estimate the exponent of
the autocorrelation function and also the exponent of the probability
distribution function. In recent years, there has been a considerable
focus on the investigation of multifractal cross-correlation between
a pair of synchronized datasets \citep{key40}.

There is a broad literature of modelling and forecasting methods of
both price and/or load time series. It usually incorportes the Autoregressive
Moving Average processes (ARMA), the Vector Autoregression (VAR),
the Vector Error Correction (VECM), machine learning, an adaptive
neuro-fuzzy network and a customers segmentation. Fixed mean, restricted
variance and normally distributed error term represent basic assumptions
for finding the best linear unbiased estimation, for a summary see
Ref \citep{key42,key45,key46}.

In this paper we study a dataset of electric power load in the Czech
Republic since $2008$ till $2011$ with a one-minute time step. We
focus on the properties of the fluctuation function where the first
periodic part of the signal is filtered from the dataset and then
the MFDFA is used. Our main aim is to determine the Hurst exponent
which provides information about the autocorrelation function as well
as the probability distribution. We also validate the assumptions
of the normal (Gaussian) noise distribution and the short-range correlations.

The paper is organized as follows: In section 2 we describe the methodology
of data processing. We first describe the Fourier filtering method
and then the MFDFA. In section 3 we analyze the dataset using the
methodology from section 2. Finally, in section 4, we draw the conclusions
of our study.

\begin{figure}
\begin{centering}
\includegraphics[scale=0.35]{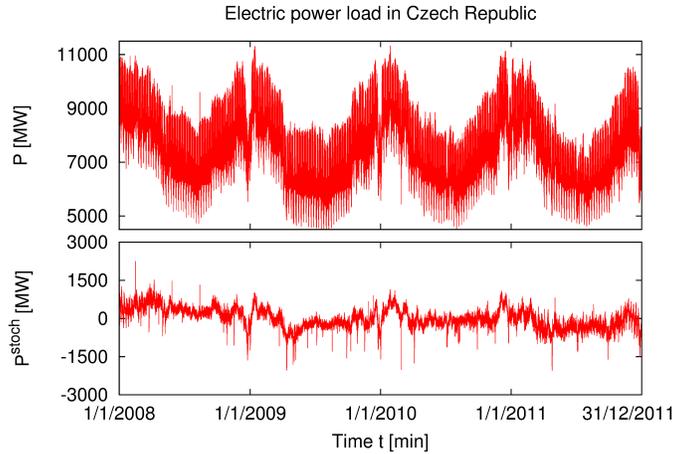} 
\par\end{centering}

\protect\protect\caption{The electric power load in Czech Republic between January 1$^{st}$
2008 and December 31$^{st}$ 2011 (top) and the stochastic part $P^{stoch}$
obtained by the filtration of the signal (bottom).}

\label{fig:Load} 
\end{figure}

\begin{figure}
\begin{centering}
\includegraphics[scale=0.35]{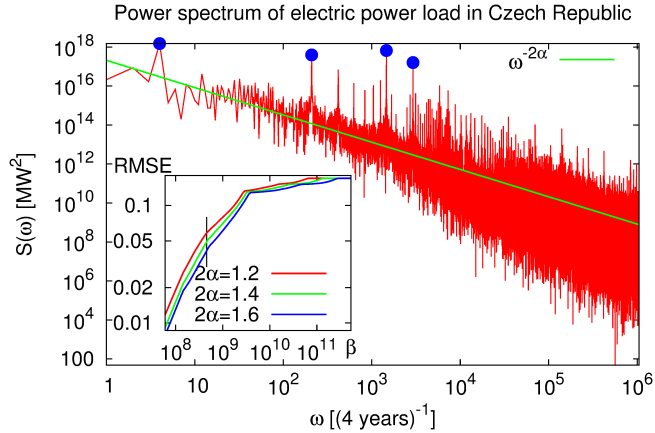} 
\par\end{centering}

\protect\protect\caption{The power spectrum of the electric power load between 1$^{st}$ of
January 2008 and 31$^{st}$ of December 2011 spanning $4$ years in
total showing the 1/f property with the extra peaks. The blue points
show one-year (Y), one-week (W), one-day (D) and 12 hour periods (12H).
In the inset of the plot, there is the dependence of the RMSE on $\beta$
for the different exponents of filtration (parameter $\alpha$). The
vertical black line shows the actual location of the parameter $\beta$
used for the filtration procedure.}

\label{fig:Power_spectrum} 
\end{figure}

\section{Methodology}

Human behavior datasets typically exhibit the oscillations with the
periods related to the units of calendar \citep{key02,key13} and
the same applies to the electric power load. The one-year and one-week
oscillations are clearly visible in Fig. \ref{fig:Load} but the presence
of other frequencies is not so easily observable. To obtain the information
regarding the strength of the oscillations we employ the Power spectrum
which is shown in Fig. \ref{fig:Power_spectrum}. It depicts additional
periods with the lengths of one day and $12$ hours beside the others.
Moreover, since the periods of the power loads do not follow harmonic
functions, we can also observe peaks at the positions of the integer
multiples of a typical trend. The reconstruction of the original load
on the basis of these most significant trend components is influenced
by randomness which is represented by less significant components
of the Power spectrum.

\subsection{Motivation}

In our study, we focus on the properties of the random part and we
use the MFDFA which is popular among scientists \citep{key16,key14,key15,key23,key25,key03,key06,key07,key11}
as an effective tool for extracting the properties of a long-range
memory within the time series.

Since time series generally might be non-stationary, polynomial trends
may still govern them. The basic idea of the DFA is to strip off the
trends and use the residues for the further analysis. In the MFDFA,
we are looking for typical patterns, which govern the time series
manifesting a self-affine property defined by $X\left(c\cdot t\right)=c^{H}\cdot X\left(t\right)$.
The generalized Hurst exponent H, determined by the method, is the
measure of the long term memory in the time series and it is directly
related to the non-integer fractal dimension D.

The disadvantage of this method is that the periodic trends disturb
the estimation of the Hurst exponent \citep{key09} and therefore,
before we employ the method, we have to filter out the oscillations
from the signal.

We use the Fourier transform to execute the filtration. The MFDFA
itself then removes the polynomial trends. The resulting signal is
decomposed as

\begin{eqnarray}
P\left(t\right) & = & P^{stoch}\left(t\right)+P^{deter}\left(t\right),\label{eq:Decomposition_of_signal}
\end{eqnarray}
where $P^{deter}$ describes the periodic behavior of the system,
while $P^{stoch}$ stands for the random part.

We used a regression model with dummy variables indicating holidays
and we perform the method described below. We observed negligible
differencies for the low orders of the MFDFA but the observable differences
for the higher orders of the MFDFA. However, the widths of the multifractal
spectrums are negligible in both cases.

\subsection{Mathematical description}

We execute our analysis in three steps. First, we perform the Fourier
transform to separate the signal into the stochastic and the deterministic
part by the Fourier transform. In the next step, we execute the MFDFA.
Finally, in the last step, we calculate the correlated Hurst exponent,
which requires shuffling of the original-time series. It is an improvement
of the typically used generalized Hurst exponent, exploited in cases,
where we have an assumption of the non-normally distributed time series.

\subsubsection{Fourier transform filtering}

We deal with a signal in the discrete time-steps $P\left(t_{n}\right)$
where $t_{n}=t_{1}+n\cdot\Delta t$ and $n\in M\equiv\{1,\ldots,N\}$.
Since the Discrete Fourier Transform of the signal is $\widehat{P}\left(m\right)=\frac{1}{\sqrt{N}}\sum_{n\in M}\exp\left(-\frac{2\pi\boldsymbol{i}\cdot n\cdot m}{M}\right)P\left(t_{n}\right)$,
and the related Power spectrum $S\left(m\right)=\widehat{P}\left(m\right)\cdot\widehat{P}^{*}\left(m\right)$,
where $x^{*}$ stands for conjugation). The Power spectrum, see Fig.
\ref{fig:Power_spectrum}, of the signal $P\left(t_{n}\right)$ exhibits
a power law-like shape with extra peaks and each coefficient of the
Fourier transform is separated into two parts according to the threshold
$\beta\cdot m^{-\alpha}$: 
\begin{itemize}
\item discrete significant coefficients in the Power spectrum for certain
frequencies above the threshold forms $\vert\widehat{P^{deter}}\left(m\right)\vert$; 
\item coefficients below the threshold forms $\widehat{P^{stoch}}\left(m\right)$
; 
\end{itemize}
where $\alpha$ and $\beta$ are the parameters set with regard to
the chosen RMSE level. We also note that if $\widehat{P^{deter}}\left(m\right)\neq0$
then we define $\arg\widehat{P^{deter}}\left(m\right)=\arg\widehat{P^{stoch}}\left(m\right)=\arg\widehat{P}\left(m\right)$.
Otherwise $\arg\widehat{P^{deter}}\left(m\right)$ is not defined.
The Fourier transform of the sub-signals $\widehat{P^{deter}}\left(m\right)$
and $\widehat{P^{stoch}}\left(m\right)$ then follow $\widehat{P}\left(m\right)=\widehat{P^{deter}}\left(m\right)+\widehat{P^{stoch}}\left(m\right)$,
which is the Fourier transform of \foreignlanguage{british}{Eq}. \ref{eq:Decomposition_of_signal}.
By executing the the inverse Fourier transform $P\left(t_{n}\right)=\frac{1}{\sqrt{N}}\sum_{n\in M}\exp\left(\frac{2\pi\boldsymbol{i}\cdot n\cdot m}{M}\right)\widehat{P}\left(m\right)$
we obtain a deterministic part $P^{deter}\left(t_{n}\right)$ from
$\widehat{P^{deter}}\left(m\right)$ . The later part $\widehat{P^{stoch}}\left(m\right)$
is transformed to $P^{stoch}\left(t\right)$.

To measure the quality of the filter we use a root mean square error,
see inset of Fig. \ref{fig:Power_spectrum}, defined as follows:

\begin{equation}
RMSE=\frac{\sqrt{N\sum_{i=1}^{N}\left(P^{deter}\left(t_{i}\right)-P\left(t_{i}\right)\right)^{2}}}{\sum_{i=1}^{N}P\left(t_{i}\right)}.\label{eq:RMSE}
\end{equation}
The level of the error was determined both to decrease the RMSE and
to prevent $P^{stoch}$ from incorporating a periodic function that
produces the artificial behavior of the fluctuation function.

\subsubsection{Multi-fractal Detrended Fluctuation Analysis}

We employ the Multi-fractal Detrended Fluctuation Analysis (MFDFA)
for analyzing the filtered signal $P^{stoch}\left(t_{i}\right)$.
The method is employed as an effective tool to avoid the artificial
\foreignlanguage{british}{behaviour} (see Ref. \citep{key22}) in
the autocorrelation function or in the Power spectrum due to the oscillation
of the electric power loads and the presence of the peaks in the Power
spectrum, see Fig. \ref{fig:Power_spectrum}.

Each element $\left\{ x_{i}\equiv P^{stoch}\left(t_{i}\right)\right\} $
of the dataset is indexed by $i\in M$. The application of the MFDFA
consists of five steps:

Step 1. Integration of the dataset to produce the dataset $X_{j}=\sum_{i=1}^{j}x_{i}$.
The ``double'' integration of the dataset $\tilde{X}_{j}=\sum_{i=1}^{j}X_{i}$
is also performed.

Step 2. Division of the dataset $X_{i}$ into $L_{s}\equiv\left\lfloor \frac{N}{t}\right\rfloor $
overlapping segments $X_{j,k}$ with length $s$ and $j\in\left\{ 1,\ldots,s\right\} $.

Step 3. Use of a standard (least-square) regression method of fixed
order $M$ on each segment $X_{j,k}$ to obtain the local trend $T_{k}\left(x\right)$
in the region $x\in\left[1,s\right]$.

Step 4. Calculation of the sample variance for each of the $L_{s}$
segments of the original dataset 
\begin{equation}
V\left(k\right)\equiv\frac{1}{s}\sum_{j=1}^{s}\left(X_{j,k}-T_{k}\left(j\right)\right)^{2}.
\end{equation}

Step 5. Averaging over all the segments of the original dataset to
obtain the multi-fractal fluctuation function

\begin{equation}
F_{q}\left(s\right)\equiv\begin{cases}
\left(\frac{1}{L_{s}}\sum_{k=1}^{L_{s}}V^{\frac{q}{2}}\left(k\right)\right)^{\frac{1}{q}} & \textrm{if}\ q\neq0\\
\exp\left(\frac{1}{2\cdot L_{s}}\sum_{k=1}^{L_{s}}\ln V\left(k\right)\right) & \textrm{if}\ q=0
\end{cases}.\label{eq:Fluctuation_function}
\end{equation}

In the analysis we investigate the properties of the fluctuation function
$F_{q}\left(s\right)$ on the window of the size $s$ and on the measure
$q$. Generally, $F_{q}\left(s\right)$ grows with increasing $s$
for all $q$ (see Fig. \ref{fig:MF-DFA-2} or follow original literature
\citep{key06,key07,key09,key15,key14,key23,key27}), following the
power law

\begin{eqnarray}
F_{q}\left(s\right) & \sim & s^{H\left(q\right)+1}.\label{eq:Definition of Hurst exponent}
\end{eqnarray}

The exponent $H\left(q\right)$ is called the Hurst exponent, see
Ref. \citep{key29}. Generally, it is related to the long-term autocorrelation
or the heavy-tailed distribution of the governing process, see Ref.
\citep{key06,key07}. We also note that $+1$ in Eq. \ref{eq:Definition of Hurst exponent}
stands due to the application of the double integration instead of
the single integration of dataset, for discussion, please, see Ref.
\citep{key06}.

We exploit a fractal spectrum to analyze whether the dataset is governed
by a single exponent or by a set of exponents. We define a scaling
function by formula:

\begin{equation}
\tau\left(q\right)=q\cdot H\left(q\right)-1.\label{eq:scaling_function}
\end{equation}
We define a fractal spectrum as the Legendre transform of $\tau\left(q\right)$
using the definition of a new variable $\pi=\frac{d\tau}{dq}$:

\begin{equation}
f\left(\pi\right)=q\cdot\pi-\tau.\label{eq:multifractal_spectrum}
\end{equation}

Generally, the fractal spectrum allows to distinguish mono- and multifractal
processes. The width of the fractal spectrum is defined by the formula
$\Delta\pi=\max_{q\in\mathbb{{R}}}\pi-\min_{q\in\mathbb{{R}}}\pi$.
The value of $\pi$ in peak of $f\left(\pi\right)$ denoted by $\pi^{max}$
represents the most frequent value of the exponent. As the width of
the fractal spectrum goes wider, the number of admitted exponents
increases and the monofractality shifts to the multifractality.

\subsubsection{Shuffling of the stochastic part of the time series}

Generally, if a stochastic process generates the time series following
a non-normal (non-Gaussian) distribution, the generalized Hurst exponent
$H\left(q\right)$ combines the information about the autocorrelation
function influenced by the properties of its probability distribution.
We extract the correlation Hurst exponent $H^{cor}\left(q\right)$
that separates the generalized Hurst exponents calculated using the
original time series and calculated using the shuffled one\footnote{To shuffle the dataset we utilized Fisher-Yates algorithm that is
effective even in the case of large dataset. In our case, we used
the average of 100 samples of shuffling.}.

While executing the shuffling procedure, we destroy the autocorrelations
(if present) within the sample. Then we use a standard MFDFA described
in previous section to the calculate shuffled fluctuation function:

\[
F_{q}^{shuf}(s)=\overline{F_{q}\left(\left\{ x_{i}\right\} ^{shuf}\right)(s)},
\]
where $\overline{x}$ stands for the averaging samples of shuffling
and $\left\{ x_{i}\right\} ^{shuf}$ means shuffling of the time serie
$x_{i}$. Finally, we estimate the generalized Hurst exponent of the
shuffled time serie $H^{shuf}\left(q\right)$. As it was noted in
the previous paragraph, the correlation Hurst exponent is then defined
by following formula:

\begin{equation}
H^{cor}(q)=H(q)-H^{shuf}(q).\label{eq:CorrelatedHurstexponent}
\end{equation}
Analogically to the generalized Hurst exponent $H\left(q\right)$
we can define the correlation fractal spectrum $f^{cor}\left(\alpha\right)$
related to $H^{cor}\left(q\right)$ by the formulas \ref{eq:scaling_function}
and \ref{eq:multifractal_spectrum}.

\subsection{Implementation of the method}

We used a multi-threaded implementation of the MFDFA with Zarja library
\citep{key01} \footnote{Source code can be found at \href{http://zarja.sourceforge.net}{http://zarja.sourceforge.net}.}
which can effectively run on multi-core cluster computers. We also
compared the results with the implementations used in \citep{key27,key26,key23}.
The filtration of dataset was executed in the Python using the NumPy
and SciPy modules \citep{key30,Key31}.

\section{Analysis of dataset}

\begin{figure}[th]
{\center

\includegraphics[scale=0.44]{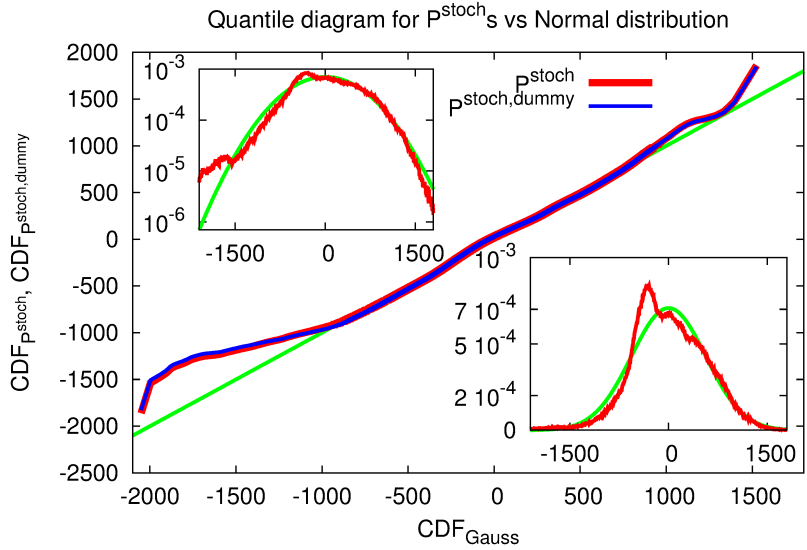}

}

\protect\protect\caption{The quantile diagram of the probability density function generated
from $P^{stoch}$ and its counterpart generated from it by the regression
model with dummy variables $P^{stoch,dummy}$ and their comparison
with the normal distribution with the same mean $\mu$ and variance
$\sigma^{2}$. In the insets, we show the comparison of the probability
density functions of $P^{stoch}$ with the appropriate normal distribution
(the normal plot at the bottom, the log-normal scale at the top).}

\label{fig:Quantile diagram} 
\end{figure}

\begin{figure}[th]
\begin{centering}
\includegraphics[scale=0.44]{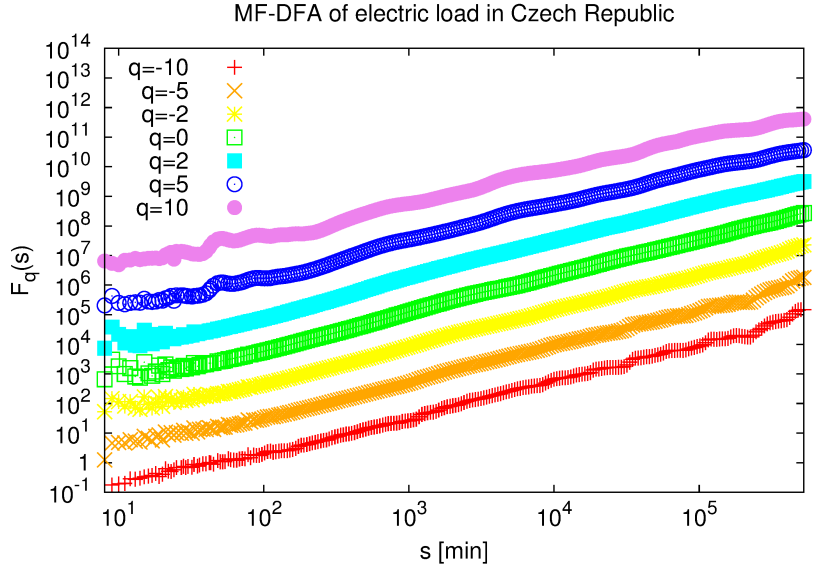} 
\par\end{centering}

\protect\protect\caption{The fluctuation function $F_{q}\left(s\right)$ of the signal $P^{stoch}$
obtained using the MFDFA of the order $4$ for various $q$s. We present
$q\in\left\{ -10,-5,-2,0,2,5,10\right\} $ from the bottom to the
top, respectively. Each plot is multiplied by factor $10$ from its
predecessor. }

\label{fig:MF-DFA-2} 
\end{figure}

\begin{figure}[th]
\begin{centering}
\includegraphics[scale=0.44]{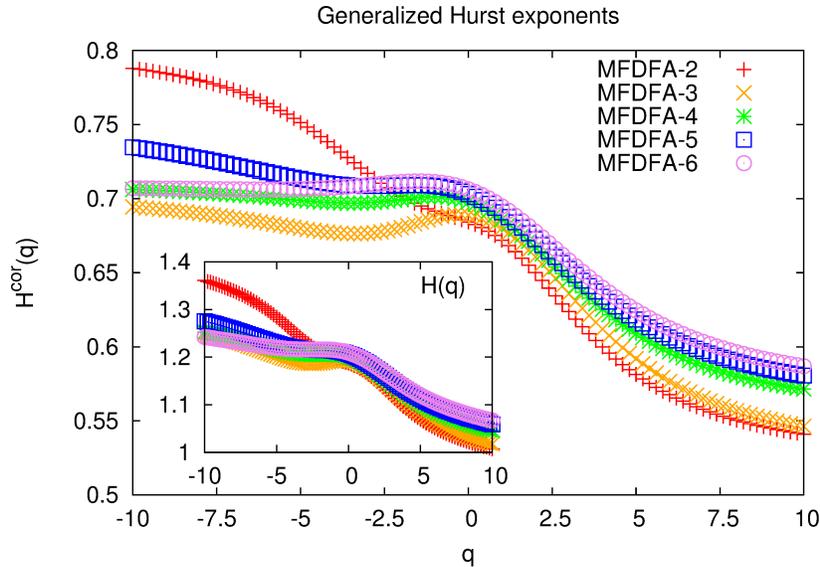} 
\par\end{centering}

\protect\protect\caption{The correlation Hurst exponent $H^{cor}\left(q\right)$ estimated
using the MFDFA of orders $2-6$. In the inset, we show the generalized
Hurst exponent $H\left(q\right)$ for the same MFDFA orders. We used
the dataset obtained by the regression model with dummy variables
indicating the holidays. The dataset without use of the method follows
the analogous pattern.}

\label{fig:Hurst exponent} 
\end{figure}

\begin{figure}[th]
{\center

\includegraphics[scale=0.45]{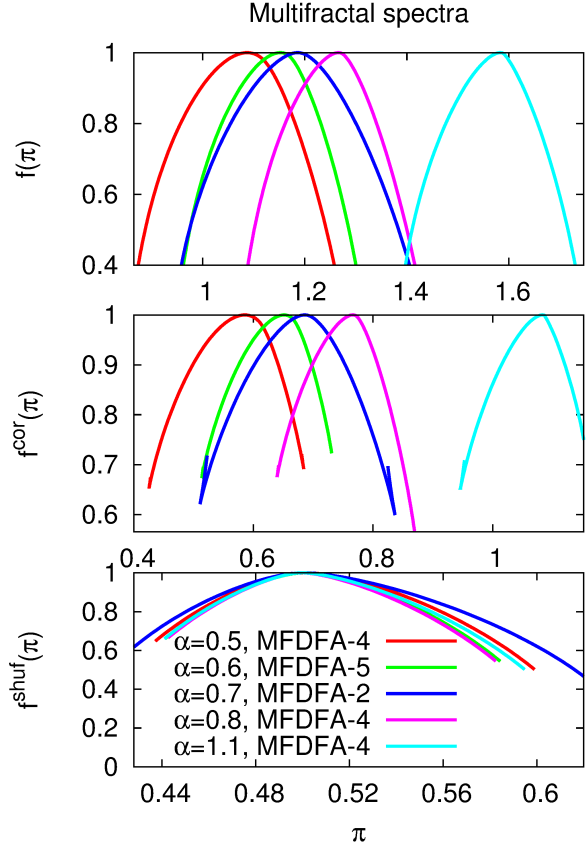}

}

\protect\protect\caption{The multifractal spectrum $f\left(\pi\right)$ of $P^{stoch}$ for
various orders of the MFDFA method and initial detreding with parameter
$\alpha$ is shown on the top. In the middle we present the correlation
and shuffled multifractal spectrum, respectively.}

\label{fig:multifractal_spectrum} 
\end{figure}

\begin{figure}[th]
{\center

\includegraphics[scale=0.38]{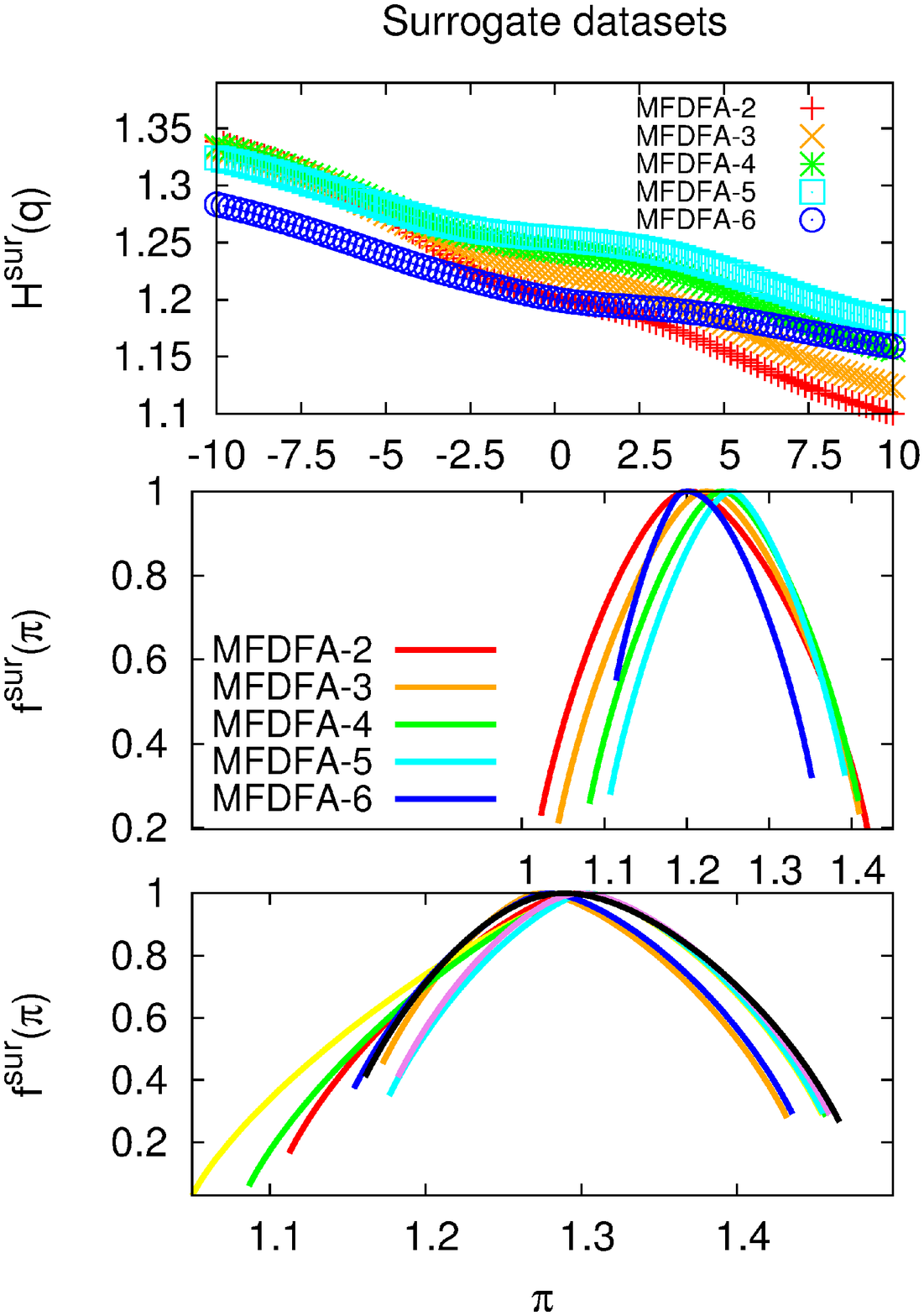}

}

\protect\protect\caption{The generalized Hurst exponents (top) and the multifractal spectrums
(middle, bottom) of the surrogate datasets which underwent phase randomization.
The middle figure show the multifractal spectrums for the surogate
dataset and bottom one illustrates dependence on the subset of the
dataset (each color shows the different subset). The top and middle
figure plots dependence on the order of the method. The bottom subplot
is for order $4$ of the method.}

\label{fig:surrogate_multifractal_spectra} 
\end{figure}

\subsection{Data description}

Our dataset describes the electric power load of the Czech Republic
which is monitored by national Transmission System Operator (TSO),
\v{C}EPS a.s. It was calculated with high frequency from the stored
data using the formula:

\begin{equation}
P\left(t\right)=\sum_{i\in1}^{M}T\left(t,i\right)-E\left(t\right)+I\left(t\right)+P_{u}\left(t\right),\label{eq:Electric power load}
\end{equation}
where $T\left(t,i\right)$ stands for $i$-th turbo-generator output
of the total number $M$. The turbo-generators are directly measured
from their minimal value of $100\ kW$ of installed capacity. $E\left(t\right)$
and $I\left(t\right)$ are the exports and imports, respectively.
Generally, they are a kind of bottlenecks because there are only few
direct transmission lines between the Czech Republic and the neighboring
countries. Finally, $P_{u}\left(t\right)$ stands for the balance
of the pumped-storage hydroelectricity\footnote{There are three of them - Dlouhé strán\v{e} $600\ MW$, Dalešice $450\ MW$
and Št\v{e}chovice with $48\ MW$ of installed capacity.}.

The dataset is calculated in real time from various sources and the
datalinks are not generally completely error-proof. Each datapoint
is thus accompanied with the confidence flag indicating the credibility
of the source. Some datapoints are calculated, using Eq. \ref{eq:Electric power load},
others are interpolated.

Our dataset consists of $N=2,103,840$ datapoints and it spans $4$
years since $2008$ till $2011$ with a one-minute time step. In our
analysis, we neglect the confidence flag and we use the electric power
load measured in $\mathrm{MW}$ only.

\subsection{Results of Fourier filtering}

The electric power load dataset of the Czech Republic is depicted
in Fig. \ref{fig:Load}, where the Power spectrum $S\left(\omega\right)$
exhibits the power law with extra significant peaks, see Fig. \ref{fig:Power_spectrum},
and therefore we first execute the Fourier filtering of the dataset
where we assume $\vert\widehat{P^{stoch}}\left(m\right)\vert=\beta\cdot m^{-\alpha}$
with parameters $\alpha$ and $\beta$ yet to be determined. In our
study we mainly choose $\text{\ensuremath{\alpha}}=0.7$ as an approximation
of the best fit of this exponent and in order to the prove robustness
of the method, we also plot the $RMSE$ for the two other values close
to the chosen value of the parameter $\alpha$, see inset of Fig.
\ref{fig:Power_spectrum}. The extensive test of the dependence of
the fractal spectra on the exponent $\alpha$ is shown in Fig. \ref{fig:multifractal_spectrum}.

Then we construct the dependence of the $RMSE$ on the parameter $\alpha$
and we choose the break-point of this dependence as an $\alpha$ value.
The $RMSE$ is defined by Eq. \ref{eq:RMSE} and at $\beta=7\cdot10^{8}\ \mathrm{MW}$
(we note that it is the equivalent of $S\left(m\right)\sim m^{-2\alpha}$).
The $P^{stoch}$ does not exhibit large periodic fluctuations (see
the bottom part of Fig. \ref{fig:Load}) and the quantile diagram
as well as the probability density distribution around the mean behave
close to the normal (Gaussian) distribution (Fig. \ref{fig:Quantile diagram}).
We note that the choice of $\beta=2\cdot10^{9}\ \mathrm{MW}$ leads
to both the significant deviation from normal distribution in its
center part as well as to the increase of the periodicity in the stochastic
part. The filtered signal is shown at the bottom of Fig. \ref{fig:Load}
and it is then more analyzed.

\subsection{Results of application of MFDFA}

Firstly, we investigate the probability distribution function of the
time series $P^{stoch}$ despite of the fact that there can still
be temporary trends, see Fig. \ref{fig:Quantile diagram}. The comparison
of the quantile diagram, the mean and the variance of $P^{stoch}$
with quantiles of the normal (Gaussian) distribution is presented
in Fig. \ref{fig:Quantile diagram}. It clearly shows the deviations
for the small values of the power load. In the lower right inset in
Fig. \ref{fig:Quantile diagram}, the comparison of the histogram
of $P^{stoch}$ with the appropriate normal distribution exhibits
a good approximation about the average. In the upper left inset in
Fig. \ref{fig:Quantile diagram}, we can observe the deviations of
the small values of the power load from the normal distribution in
the semi-logarithmic scale.

In the next step, we perform the MFDFA to calculate the fluctuation
function $F_{q}\left(s\right)$ and we estimate the generalized Hurst
exponent $H\left(q\right)$, see the inset of Fig. \ref{fig:Hurst exponent}
in range $[2\cdot10^{3},2\cdot10^{5}]$. The generalized Hurst exponent
depends on $q$ we expect presence of multifractality. To get valuable
information about the autocorrelation function, we shuffle the dataset
to calculate the fluctuation function $F_{q}^{shuf}\left(s\right)$.
The ratio of the original fluctuation function $F_{q}\left(s\right)$
against the fluctuation function of the shuffled dataset $F_{q}^{shuf}\left(s\right)$
formulated as $F_{q}^{cor}\left(s\right)=\frac{F_{q}\left(s\right)}{F_{q}^{shuf}\left(s\right)}$
follows the power law similarily as $F_{q}\left(s\right)$ see Fig.
\ref{fig:MF-DFA-2}.  Then the calculation of the correlation Hurst
exponent $H_{q}^{cor}\left(s\right)$ is performed using the formula
\ref{eq:CorrelatedHurstexponent}. We show $H_{q}^{cor}\left(s\right)$
in Fig. \ref{fig:Hurst exponent} and the exponent stands between
the values of $0.55$ till $0.8$ (in contradiction to the calculation
of the generalized Hurst exponent based on the normally distributed
time series), showing a strong persistence. Additionally we note that
the estimation of the Hurst exponents is stable with regard to the
orders of the MFDFA.

In Fig. \ref{fig:multifractal_spectrum}, the fractal spectrum $f\left(\pi\right)$,
the correlation fractal spectrum $f^{cor}\left(\pi\right)$ and also
the shuffled fractal spectrum $f^{shuf}\left(\pi\right)$ of the stochastic
part $P^{stoch}$ are not concentrated at single $\pi$ but they are
broadly spread among the wide range of $\pi$s. 
conclude that the processes are multifractal in the distribution as
well as in the correlation function. However, multifractality of the
correlation function is stronger $\Delta\pi^{cor}\cong0.3$ in contrast
to the multifractality of the distribution function $\Delta\pi^{shuf}=0.15$
for the same order of the method. 

\subsection{Tests of stability of the results}

The above mentioned results of the analysis may depend on additional
factors. To address the factors we execute additional tests to show
the invariance of the conclusions.

\subsubsection{Stability of results with respect to the filter}

As a test of the stability of the results, we performed multiple calculations
of MFDFA for different values of the parameters $\alpha$ and $\beta$.
The generalized Hurst exponent as well as the fractal spectrum depend
on a particular value of $\alpha$ and it is independent on the order
of the method, see Fig. \ref{fig:multifractal_spectrum}. The change
of the order does not significantly imply the change of the width
of the fractal spectra. On the other hand, the shuffled fractal spectrum
is independent on the value of $\alpha$ and it is localized around
$\frac{1}{2}$ -- the value of the Gaussian distribution. The persistence
of the time series is conserved in the proximity of $\alpha=0.7$,
see the middle of the Fig. \ref{fig:multifractal_spectrum}.

\subsubsection{Surrogate data test}

Generally, there are usually two reasons of the multifractality in
time series: 
\begin{itemize}
\item long range correlations of small and large fluctuations within the
time serie; 
\item heavy-tailed probability distribution function (not necessarily the
Lévy $\alpha$-stable distribution, see Ref. \citep{key55}). 
\end{itemize}
The long-range correlation property and the fat-tailed probability
distribution are investigated by shuffling and by a phase randomization.
Shuffling destroys the correlations within the time series but it
preserves the probability distribution. On the other hand, the phase
randomization preserves the correlation function but weakens both
the non-Gaussian and non-linear properties of the time serie. The
procedures were firstly proposed in Ref. \citep{key50} and a review
of its use can be found in Ref. \citep{key51}. We note that this
method was initially used in the context of the MFDFA in Ref. \citep{key52}.

We practically performed the test on $50$ samples of the surrogate
datasets and we present the results in Fig. \ref{fig:surrogate_multifractal_spectra}.
In the graph in the top we can see similar results of $H$ as in the
inset in the Fig. \ref{fig:Hurst exponent}. In the middle graphs
there is the result comparable with the top graphs in the Fig. \ref{fig:multifractal_spectrum}.
We obtained the width of the fractal spectra $\Delta\pi\cong0.3$
and the location of the maximum is around $\pi^{max}\cong1.2$. We
conclude that the multifractality is not caused by non-linearity and
beside that there are the indications of the presence of a distribution
with the tail heavier than the Gaussian distribution possess. From
theory of the stable distributions and the stochastic processes, Refs.
\citep{key55,key60}, the Gaussian distribution possess $H\left(2\right)=\frac{1}{2}$
and the Lévy $\alpha$-stable distribution $H\left(2\right)=\frac{1}{\omega}$
where $\omega$ is the exponent of the tail (for the Gaussian distribution
we have $\omega=2$). We obtained for the shuffled multifractal spectra,
where shuffling erases the autocorrelations with in the time series,
see the bottom of Fig. \ref{fig:multifractal_spectrum}, wide peak
around $\pi^{shuf}=\frac{1}{2}$. Based on the assumption that the
probability distribution is stable we admit presence the Lévy $\alpha$-stable
distribution with the exponents $\omega$ close to the values of the
Gaussian distribution. We also note that the result is independent
of the set up of the initial filtering method. Additionally the Lévy
$\alpha$-stable distribution must be skewed due to indications in
Fig. . We also tested the influence of using a regression model with dummy variables for the decrease of 
the effect of holidays. As you can see on Fig. \ref{fig:Quantile diagram}, the result is not 
significant.


\subsubsection{Problems of stationarity and deficient random generators}

We applied the Augmented Durbin-Watson test on the $P_{stoch}$ and
we rejected the null hypothesis of non-stationarity at the $5\%$
significance level.

As a test of the stability of the results we separated the original
dataset into $8$ sets with equal size and we executed the proposed
method for each segment. The results of the method are at the bottom
of Fig. \ref{fig:surrogate_multifractal_spectra} where the curves
representing the surrogate fractal spectra show the overlap with the
width of fractal spectra $\Delta\pi^{sur}\cong0.3$ and $\pi^{max}\cong1.3$.
These values are approximately equal to the results of the complete
dataset. Thus, we conclude, that the results of the method are stable
with the respect to the change of the scale.

\section{Conclusions and Outlook}

The main contribution of this paper is an analysis of the high-frequency
electric power loads dataset of the Czech Republic using the improved
MFDFA methodology. We discovered that the power spectrum of the signal
exhibits 1/f noise property with the additional peaks that are caused
by a periodic behavior of the electricity consumption. Based on that
fact, we first separated the noise from modulating signal and then
we applied the MFDFA without dealing with an artificial behavior of
the fluctuation function, see Ref. \citep{key09}. After that we exploited
the MFDFA for the analysis of the dataset to obtain information about
the autocorrelation function. The major part of the power load is
governed by oscillations. Beside that we report a strong persistence
of the power loads where the distribution function exhibits non-Gaussian
properties. The fractal spectra of both the distribution as well as
the autocorrelation function indicate the presence of multifractality.
We also performed a test using surrogate datasets as well as a test
of stationarity to validate the strength of our conclusions. The analysis
suggests the presence of the probability distribution with the tails
heavier than the Gaussian distribution.

Some of our results are in contradiction with the previously published
work analyzing electricity consumption and also with the assumptions
of electricity load prediction models \citep{NowickaZagrajek20021903,key02,Pappas20081353,Pappas2010256}.
First, our analysis indicates that the stochastic part of the signal
is not normally distributed, second, the distribution function is
skewed and it may even have infinite moments of the probability distribution
and third, the autocorrelation function is persistent. We also conclude
that the estimations of risks based on traditional forecasting methods
using the Gaussian distribution and short-range correlations are not
usable due to both the long-range autocorrelation and the probability
distribution's extremes. The main part of the load constituting approximately
$95\%$ of the signal was filtered out and it is systematically driven
by external factors. Modeling by means of a regime-switching model
makes a good sense to us.

The Czech transmission system is sufficiently dimensioned to cope
with electricity consumption fluctuations contained in the dataset
we had at our disposal. The problem that attracts actual attention
of the TSO is dealing with the unexpected flows from north to south
of Europe through the Czech Republic, see Ref. \citep{key28}, which
are caused by inhomogeneity of sources generating electricity and
consumption of electricity in Europe. The presented approach might
also be applied to solve a more complex problem, where in addition
to the uncertainty of the electricity consumption, we may also consider
the uncertainty caused by real power inflows and outflows (imported
and exported electricity) or the uncertainty due to differences between
cross-border trading and real electricity flows (obeying Kirchhoff's
laws). The level of uncertainty is expressed as a deviation from foreseeable
behavior described by polynomial trends and periodic oscillations.

\section*{Author contributions}

J.K. obtained and prepared the dataset. H.L. prepared the tool for
analysis and performed the analysis. J.K. and H.L. contributed to
the writing of the manuscript. The work described in this paper will
be used in J.K.'s Ph.D. thesis.

\section*{Acknowledgement}

 This article was supported by Czech Ministry of Education RVO68407700 and it also was written with the support of SVV project Strengthenning Doctoral Research in Economics and Finance. 
  We thank for the fruitful discussion to P. Jizba, J. Lavi\v{c}ka,
A.M. Povolotsky, V.B. Priezzhev, E. Lutz, T. Kiss, G. Alber and H.E.Stanley.

\section*{Parameters and symbols of the methodology}

{\center

\begin{tabular}{|c|c|c|}
\hline 
Value  & Symbol  & Unit\tabularnewline
\hline 
\hline 
Electric load  & $P$  & $\mathrm{MW}$\tabularnewline
\hline 
Stochastic part of electric load  & $P^{stoch}$  & $\mathrm{MW}$\tabularnewline
\hline 
Window size  & $s$  & $\mathrm{min}$\tabularnewline
\hline 
Multifractal measure (parameter)  & $q$  & $1$\tabularnewline
\hline 
Multifractal fluctuation function  & $F_{q}\left(s\right)$  & $\mathrm{MW}$\tabularnewline
\hline 
Generalized Hurst exponent  & $H\left(q\right)$  & $1$\tabularnewline
\hline 
Hurst exponent  & $H\equiv H\left(2\right)$  & $1$\tabularnewline
\hline 
Correlation Hurst exponent  & $H^{cor}\left(q\right)$  & $1$\tabularnewline
\hline 
Hurst exponent of shuffled time serie  & $H^{shuf}\left(q\right)$  & $1$\tabularnewline
\hline 
Scaling exponent  & $\tau\left(q\right)$  & $1$\tabularnewline
\hline 
\end{tabular}

}

\bibliographystyle{elsarticle-num}

\end{document}